\documentclass{article}
\pdfoutput=1
\PassOptionsToPackage{numbers, compress}{natbib}

\usepackage[preprint]{neurips_2020}

\usepackage[utf8]{inputenc} 
\usepackage[T1]{fontenc}    
\usepackage{hyperref}       
\usepackage{url}            
\usepackage{booktabs}       
\usepackage{amsfonts}       
\usepackage{nicefrac}       
\usepackage{microtype}      
\usepackage{amsmath}
\usepackage{graphicx}
\usepackage{subcaption}
\usepackage{authblk}

\newcommand\headercell[1]{%
   \smash[b]{\begin{tabular}[t]{@{}c@{}} #1 \end{tabular}}}
\usepackage[]{algorithm2e}
\makeatletter
\renewcommand{\@algocf@capt@plain}{above}
\makeatother
\usepackage{float}
\floatstyle{plaintop}
\restylefloat{table}

\title{Smile-GANs: Semi-supervised clustering via GANs for dissecting brain disease heterogeneity from medical images}

\author{Zhijian Yang}
\author{Junhao Wen}
\author{Christos Davatzikos}
\affil{Center for Biomedical Image Computing and Analytics,
  Perelman School of Medicine,
  University of Pennsylvania,
  Philadelphia, PA 19104, USA}

\begin{document}

\maketitle

\begin{abstract}
Machine learning methods applied to complex biomedical data has enabled the construction of disease signatures of diagnostic/prognostic value. However, less attention has been given to understanding disease heterogeneity. Semi-supervised clustering methods can address this problem by estimating multiple transformations from a (e.g. healthy) control (CN) group to a patient (PT) group, seeking to capture the heterogeneity of underlying pathlogic processes. Herein, we propose a novel method, Smile-GANs (SeMi-supervIsed cLustEring via GANs), for semi-supervised clustering, and apply it to brain MRI scans. Smile-GANs first learns multiple distinct mappings by generating PT from CN, with each mapping characterizing one relatively distinct pathological pattern. Moreover, a clustering model is trained interactively with mapping functions to assign PT into corresponding subtype memberships. Using relaxed assumptions on PT/CN data distribution and imposing mapping non-linearity, Smile-GANs captures heterogeneous differences in distribution between the CN and PT domains. We first validate Smile-GANs using simulated data, subsequently on real data, by demonstrating its potential in characterizing heterogeneity in Alzheimer's Disease (AD) and its prodromal phases. The model was first trained using baseline MRIs from the ADNI2 database and then applied to longitudinal data from ADNI1 and BLSA. Four robust subtypes with distinct neuroanatomical patterns were discovered: 1) normal brain, 2) diffuse atrophy atypical of AD, 3) focal medial temporal lobe atrophy, 4) typical-AD. Further longitudinal analyses discover two distinct progressive pathways from prodromal to full AD: i) subtypes 1 - 2 - 4, and ii) subtypes 1 - 3 - 4. Although demonstrated on an important biomedical problem, Smile-GANs is general and can find application in many biomedical and other domains. 
\end{abstract}

\section{Introduction}
Numerous studies have been conducted via case-control group comparisons for neuroimaging biomarkers tracking in brain diseases, such as Alzheimer's Disease (AD)\cite{ad_biomarker1}\cite{ad_biomarker2}\cite{ad_biomarker3}. However, those studies suffer from underpowered statistical inferences due to the violation of the underlying assumption that each group population are relatively homogeneous pathologically, thus drawing inconsistent conclusions across studies. \par 
Better understanding of brain disease heterogeneity paves the road for precision diagnostics, as umbrella disease classification can be broken down to more precisely and homogeneously defined pathologies. Machine learning (ML) has shown great promise in this area. Semi-supervised clustering methods were recently proposed to address this issue\cite{HYDRA} \cite{Chimera}. Instead of clustering directly in patient population based on similarity/dissimilarity, semi-supervised methods seek to cluster by multiple transformations or patterns between the subgroup of patients and a reference group (e.g., cognitive normal control (CN) group). This way, they attempt to avoid capturing uninformative variations due to various confounds, and focus on variation of disease effects.  Nevertheless, their limitations lie in that the clustering either relies on the SVM-based classification accuracy \cite{HYDRA} or has strong assumptions on data distribution and transformation linearity \cite{Chimera}. More recently, deep learning (DL) has made a big leap in medical imaging applications \cite{dl_med}. Generative adversarial networks (GANs) are well-known for modeling a distribution from samples \cite{GAN}. This attribute makes it natural for learning difference in distribution of two groups of data and drives the current work to explore its potential for semi-supervised clustering. \par
To address the aforementioned limitations, we proposed a novel method, Smile-GANs: SeMi-supervIsed cLustEring via GANs, for parsing disease heterogeneity. Smile-GANs aims to tackle heterogeneity by transforming data from CN domain $X$ to patient (PT) domain $Y$. The first novelty here is to learn several distinct mappings $f_i$ such that the distributions of $f_i(X)$ are not distinguishable from the distribution of $Y$. By focusing on the difference between CN and subpopulations/subtypes in PT domain, each mapping can represent a unique neuroanatomical pattern related to disease effects. For that purpose, Smile-GANs borrowed the idea of Cycle-GAN \cite{cycle_gan} and Cluster-GAN \cite{clustergan}, constructing one-to-many mappings from CN to PT domain with unpaired data. Moreover, those multiple mappings offer interpretable neuroanatomical patterns for each subtype. The second novelty of Smile-GANs is the interactive training of the mapping function and a clustering function which transform Fake-PT domain back to the subtype domain, allowing for quick and accurate clustering of unseen PT data. The third novelty is the construction of effective monitoring criteria for training of Smile-GANs on the lower dimensional representation of imaging data, guaranteeing the performance of saved models in mapping and clustering.\par
We first validate the potential of Smile-GANs on simulated data with the known number of clusters/subtypes (\emph{K}) and their atrophy patterns. We demonstrate that Smile-GANs both accurately clusters pre-selected subtypes and discovers their corresponding simulated atrophy patterns. We then apply Smile-GANs to Alzheimer's Disease Neuroimaging Initiative (ADNI) 2 baseline data and reveal four reproducible subtypes in AD and mild cognitive impairment (MCI) subjects with distinct clinical profiles. Further analyses indicate two different disease progressive pathways by applying the trained model to ADNI 1 and Baltimore Longitudinal Study of Aging (BLSA) longitudinal data. 

\section{Method}
\begin{figure}
    \centering
    \includegraphics[width=0.8\linewidth,height=5.95cm]{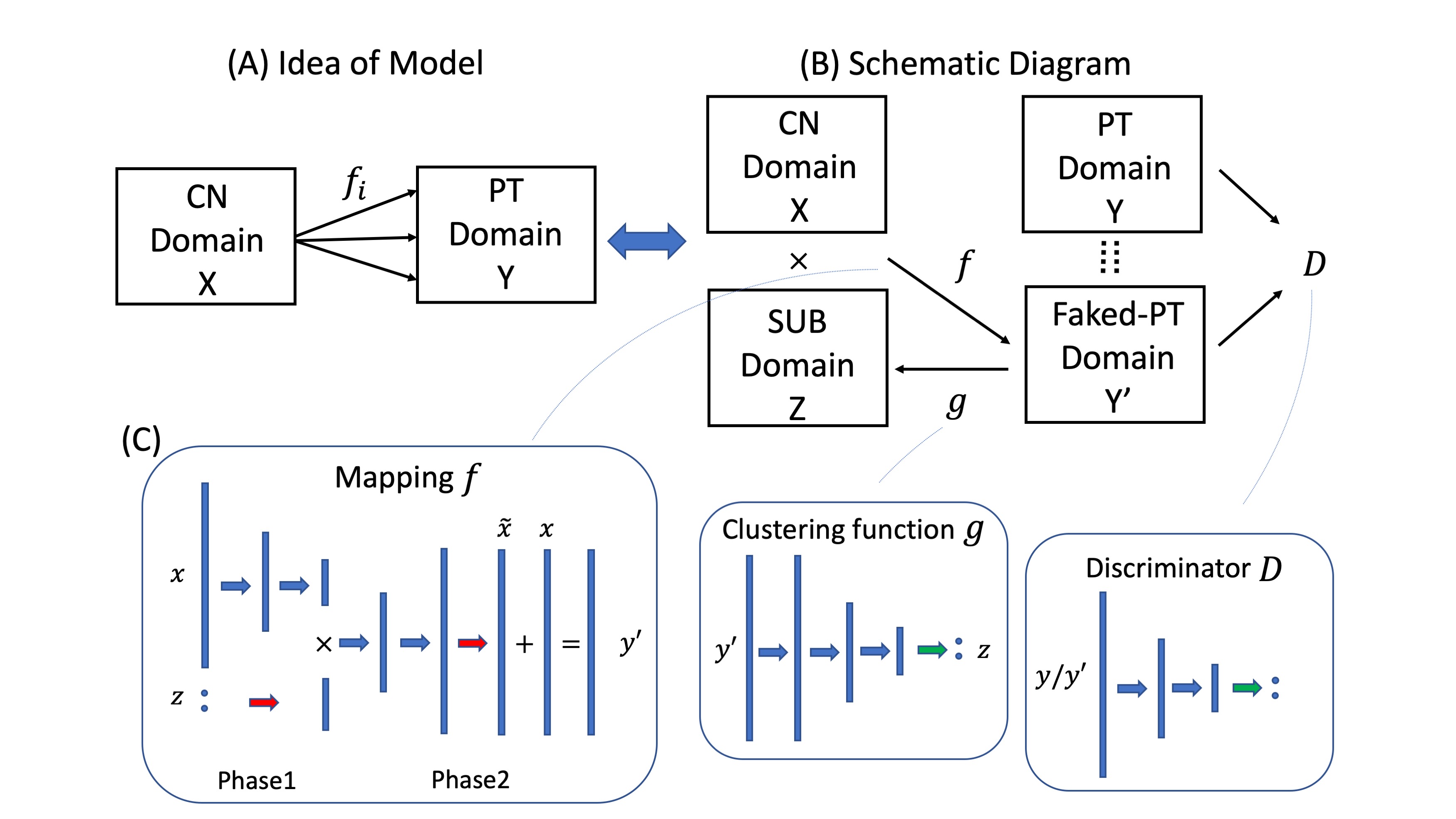}
    \caption{Schematic diagram and network architectures. (A): General idea behind Smile-GANs (B): Schematic diagram of Smile-GANs. CN: cognitive normal control, PT: patient, SUB: subtype (C): Network architecture of three functions: blue arrow represents one linear transformation followed by one leaky relu function, green arrow represents one linear transformation followed by one softmax function, red arrow represents only one linear transformation.} \label{fig1}
\end{figure}
The general structure of Smile-GANs is shown in Fig.~\ref{fig1}. The essential element of the model is to learn one-to-many mappings from CN domain $X$ to PT domain $Y$. The idea is equivalent to learning one mapping function $f$ which generates fake PT data $y'=f(x,z)$ from joint domain $X$ $\times$ $Z$, while enforcing the equality between PT domain $Y$ and Fake-PT domain $Y'$. Here, $Z$ is referred as the subtype (SUB) domain and we denote data distributions in four domains as $x\sim p_\text{CN}(x)$, $y \sim p_\text{PT}(y)$, $y' \sim p_{Y'}$and $z \sim p_\text{SUB}(z)$, respectively. The variable $z\in Z$, independent from $x$, can take a value from 1 to \emph{K} (i.e., the number of subtypes/mappings) with equal probability. In addition, an adversarial discriminator $D$ is introduced to distinguish between real PT $y$ and fake PT $y'$. On top of that, we introduce another function $g$ from domain $Y'$ to $Z$ which serves both as a regularization term and a clustering function from PT domain to SUB domain for clustering membership assignment.\par 
The objective contains two types of losses. First, the adversarial loss \cite{GAN} serves to match the distribution of data in Fake-PT domain to real data in PT domain. Secondly, the regularization loss includes the change loss and the cluster loss. Specifically, the change loss controls the distance of transformation, with the assumption that disease effects should not greatly change the original anatomy. The cluster loss controls the independence among multiple mappings and guarantees the clustering accuracy of function $g$ on patient data. We give more details of the objective in the following sections.\par

\subsection{Adversarial Loss}\label{section2.1} 
The adversarial loss is applied for training of discriminator $D$ and the mapping function $f$, which can be written as:
\begin{align*}
    L_\text{GAN}(D,f)&=E_{y\sim p_\text{PT}}[log(D(y))]+E_{z\sim p_\text{SUB},x\sim p_\text{CN}}[log(D(f(x,z)))]\\
    &=E_{y\sim p_\text{PT}}[log(D(y))]+E_{y'\sim p_\text{Y'}}[log(D(y'))]
\end{align*}\par
 The mapping $f$ attempts to transform CN to corresponding fake PT data so that they follow similar distributions as real PT data. The discriminator $D$, representing the probability that $y$ come from the real data rather than generator, is trying to identify the fake PT data from the real PT data. Therefore, the discriminator attempts to maximize the adversarial loss function while the 
mapping $f$ attempts to minimize against it. The training process can be denoted as:
\begin{align*}
    \min_{f}\max_{D}L_\text{GAN}(D,f)=E_{y\sim p_\text{PT}}[log(D(y))]+E_{y'\sim p_\text{Y'}}[log(D(y'))]
\end{align*}\par
\subsection{Regularization Loss}\label{section2.2}
The change loss is to control the distance of transformations. As our model is applied to a lower dimensional representation of imaging data, region of interests (ROIs), we assume that only some specific regions will be affected by the disease process, whereas the rest should remain unchanged. To encourage sparsity, we define the change loss to be the $l_1$ distance between the fake PT data and the original CN data: $L_\text{change}(f)=E_{x\sim p_\text{CN},z\sim p_\text{SUB}}[||f(x,z)-x||_1]$.\par
Moreover, we formulate the cluster loss as $L_\text{cluster}(f,g)=E_{x\sim p_\text{CN},z\sim p_\text{SUB}}[||g(f(x,z))-z||_2]$. By controlling the distance between the sampled SUB variable z and the reconstructed SUB variable, we enforce the function $g\circ f$ to be an Identity function of $z$. This leads to the first property that $f(a,z)$ is an injective mapping for fixed $x=a$ such that different values of SUB variable $z$ will transform the same CN data to different PT data. With minimization of the cluster loss, there arises the second property: for $z_1\neq z_2$ and different CN data, $x_1\neq x_2$ $f(x_1,z_1)\neq f(x_2,z_2)$. These two properties of the mapping function $f$ are important, since they guarantee that the SUB variable $z$ is not ignored during training process and that there is no intersection among mapping directions (i.e., one PT data is only assigned to one subtype).\par
More importantly, with the assumption that $p_\text{PT}=p_\text{Y'}$ after $f$ and $g$ are properly trained \cite{GAN}, we can derive the equality between PT domain $Y$ and Fake-PT domain $Y'$ and thus for $y\sim p_\text{PT}$, $g(y)=g(y')=g(f(x,z))=z$, where $z\sim p_\text{SUB}$ indicates the subtype membership. In other words, $g$ can be a clustering function for quickly clustering unseen data.

\subsection{Full Objective}
With the aforementioned losses, we can write the full objective as:
\begin{align*}
L(D,f,g)=L_\text{GAN}(D,f)+\mu L_\text{change}(f)+\lambda L_\text{cluster}(f,g)
\end{align*}
with $\mu$ and $\lambda$ be two parameters controlling the relative importance of each Loss function during the training process. Through this objective, we want to find the mapping f and the clustering function g such that:
\begin{align*}
f,g=\arg \min_{f,g} \max_{D} L(D,f,g)
\end{align*}

\section{Implementation details}
\subsection{Network Architecture}
For faster convergence of model in implementation, the mapping function, instead of directly transforming the CN data to the fake PT data, first learns a change in the CN data and takes the sum of them to obtain the fake PT data.
Therefore, the architecture of the mapping function $f$ can be divided into two phases as shown in Fig.~\ref{fig1}(C). In the first phase, the CN data and the SUB variable are mapped to latent representations with same dimension through encoder and decoder \cite{autoencoder}, respectively. The second phase has one decoding structure mapping the dot-product of two representations to the change $\tilde{x}$, which is added to the CN data $x$ to acquire the fake PT data. 
The discriminator $D$ and the clustering function $g$ have similar encoding structures, with $D$ mapping PT/fake PT data to prediction vector with dimension 2 while the encoder $g$ mapping the fake PT data to a SUB representation. 
\subsection{Training Details}
First, we rewrite the procedure in section \ref{section2.1} as:
$\min_D L_\text{GAN}(D)=E_{y\sim p_\text{PT}}[(D(y)-1)^2]+E_{z\sim p_\text{SUB},x\sim p_\text{CN}}[D(f(x,z))^2]$ and $\min_f L_\text{GAN}(f)=E_{z\sim p_\text{SUB},x\sim p_\text{CN}}[D(f(x,z)-1)^2]$. Using this least square loss instead of original log likelihood boosts stability of the training process \cite{LSGAN}. Second, the SUB variable z is constructed as a one-hot latent variable with dimension \emph{K} instead of one single value and thus the cross-entrophy loss is computed for the cluster loss defined in section \ref{section2.2}.\par
We set two parameters to be $\mu=5$ and $\lambda=9$ for all experiments. Also,we performed gradient clip for each iteration to avoid the explosion of gradient during the training process. For optimization, we used ADAM optimizer \cite{ADAM} with learning rate 0.0004 for Discriminator $D$ and 0.002 for mapping $f$ and clustering function $g$. $\beta_1$ and $\beta_2$ are 0.5 and 0.999, respectively. More details about architectures and training procedures are present in Supplementary.\par

\subsection{Stopping Criteria}
For real application, since the ground truth of subtypes is unknown, we adopt an approximation of the Wasserstein distance (WD) \cite{WGAN} as one metric for monitoring the training process and choosing the stopping point. For Smile-GANs, instead of deriving WD from optimization as the original paper introduced, we used the closed form formula to compute the distance. For stopping criteria, we assume that, in CN domain and in all subpopulations of PT domain, the lower dimensional representation of each data point (ROIs) is sampled from a multivariate Gaussian distribution. Though this assumption might be strong, it does enable us to estimate the WD quickly. \par
To be more specific, for each mapping direction $z=i$ and for all samples in CN domain $X=\{x_1,x_2,\cdots,x_n\}$, we calculate the mean vector $m_1^i$ and covariance matrix $C_1^i$ of $f(X,i)$. 
Also, from samples in PT domain Y, we take out the subset $Y_i=\{y_j\}^i$ such that $g(y_j)=i$ and calculate the mean vector $m_2^i$ and covariance matrix $C_2^i$. With mean vectors and covariance matrices, we can compute the 2nd Wasserstein distance using the formula for two multivaraite gaussian measure:
\begin{align*}
    W_2(\mu_{f(X,i)},\mu_{Y_i})=||m_1^i-m_2^i||^2_2+\text{trace}(C_1^i+C_2^i-2({C_2^i}^{1/2}{C_1}^i{C_2^i}^{1/2})^{1/2})
\end{align*}

If we further assume that all features are independent, we can derive diagonal covariance matrices which make the computation even faster and, based on our experiments, this assumption does not affect monitoring training process. \par
Moreover, to deal with rare cases when inconsistencies exist between WD and model performance, we also derive two other metrics: alteration quantity (AQ), which represents the number of subjects whose subtype memberships alter in the last five epochs. A small AQ represents high stability of the model. Lastly, the cluster loss, indicating the performance of clustering function $g$, is also considered as part of the stopping criteria.

\section{Experiments}
\subsection{Experiments on Simulated Data}
\begin{figure}[!tb]
  \centering
  \includegraphics[width=1\linewidth,height=8.8cm]{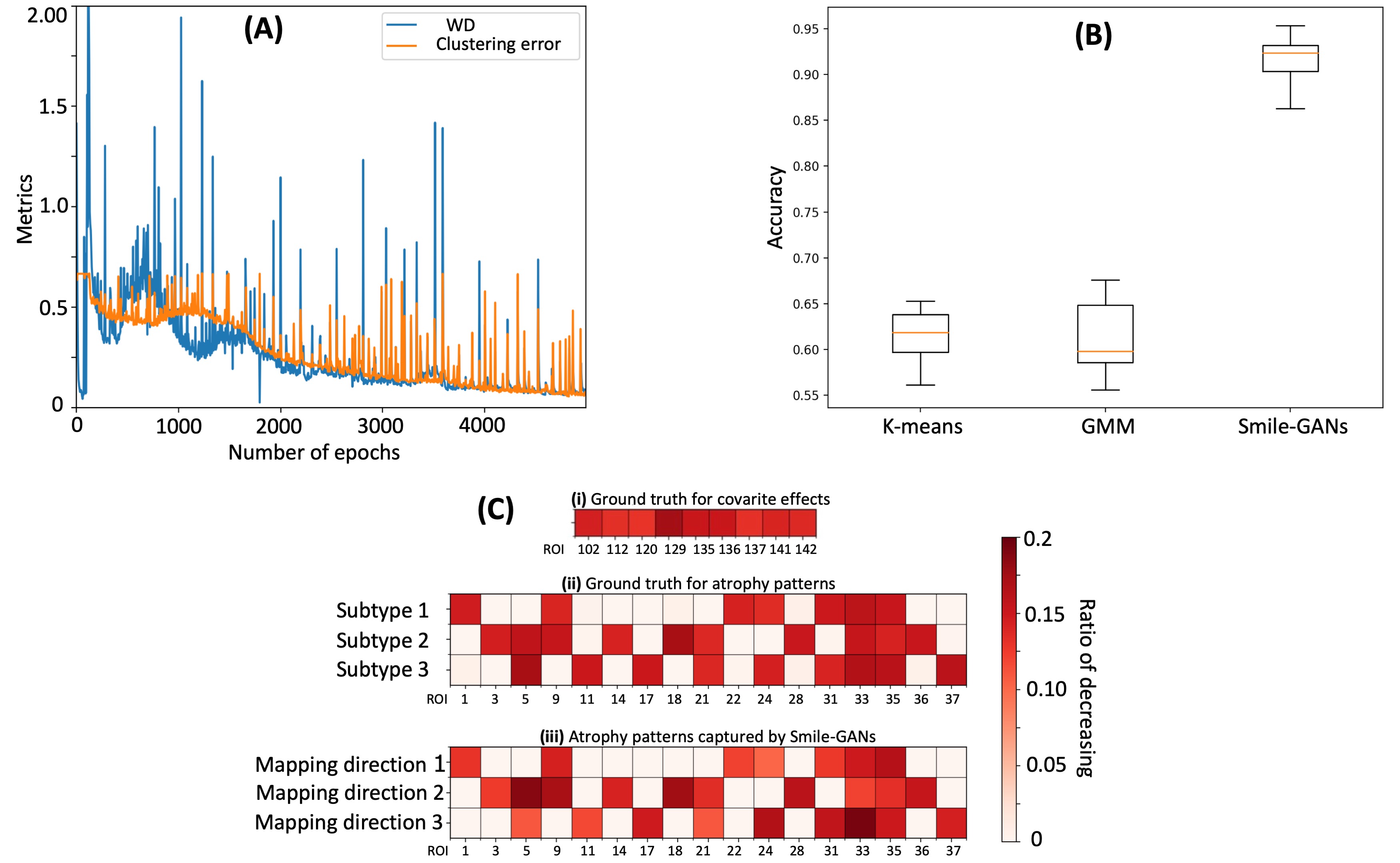}
  \caption{Results on simulated data. (A): Clustering training process monitoring with Wasserstein distance (WD). Clustering error = 1 - clustering accuracy. (B): Comparison of clustering performance across Smile-GANs, K-means and Gaussian mixture model (GMM). (C): The mapping function (\emph{f}) of Smile-GANs discovers (iii) the ground truth of subtypes' pure atrophy patterns (ii) while not confounded by covariate patterns (i). Each mapping direction represents one subtype. Only the ROIs whose values decreasing over 5\% are displayed.}\label{fig3}
\end{figure}

\textbf{Simulated Data Generation} Simulated data were generated in a low dimensional space (i.e., 145 ROIs). For each subject, the 145 ROIs were simulated by sampling from a normal distribution $\mathcal{N}(1,0.1)$. In total, 1200 subjects were generated independently and then randomly split into two half-split sets, with each (600) being CN and pseudo-PT group, respectively. The atrophy simulation was only introduced for pseudo-PT subjects, which were further divided into 3 subtypes with the same number of subjects (200). For each subtype, the values of specific pre-selected ROIs were deceased by 10 to 20\% randomly to simulate the severity of the atrophy. Moreover, to simulate covariate effects, we randomly sampled 200 subjects from both CN and pseudo-PT respectively and decreased the values by 10 to 20\% in some other ROIs. The simulation ground truth for the covariate patterns and the 3 subtypes are shown in Fig. ~\ref{fig3}(c)(i) and (ii), respectively. Note that overlapping of ROIs across subtypes was imposed to better follow the nature of atrophy. \par

\textbf{Experiments on Simulated Data} 
To add variability of the clustering performance, we repeated the simulated data generation and ran the experiment independently 20 times. We first investigated the potential of WD for monitoring training process. Then we compared the clustering performance of Smile-GANs with traditional K-means \cite{kmeans} and Gaussian Mixtrue Model (GMM) \cite{gmm}. Finally, the mapping function ($f$) of Smile-GANs can be used to visualize the subtypes' atrophy patterns captured for fake PT data generation. We calculated the mean difference between CN and fake PT in \emph{K} mapping directions and inspected ROIs with significant decreasing in values.

\subsection{Experiments on Real Data}
\textbf{Data and Image Processing} For experiments on real data, we first included 297 CN and 602 AD/MCI subjects for baseline T1-weighted (T1) MRIs from ADNI 2 database. The trained model was further applied to longitudinal data, undergoing more than one visit in 2 to 23 years from the baseline, from ADNI1 and BLSA dataset. The longitudinal data consist of 1323 CN and 610 MCI/AD subjects. For all T1 MRIs, brain tissue segmentation was performed using a multi-atlas segmentation technique \cite{MUSE} and 145 ROIs were derived as features for Smile-GANs. Those features were first harmonized to remove site effect \cite{Harmonization}, and then age and gender effects were corrected in a pooled sample of matched controls using a voxel-wise linear model. Moreover, gray matter (GM) tissue maps \cite{RAVEN} were also segmented for voxel-wise statistical mapping.\par

\textbf{Experiments on baseline data}
We first chose the optimal \emph{K}. Smile-GANs was run for 20 times for different \emph{K} (\emph{K} = 2 to 5). The optimal \emph{K} was chosen by the highest Adjusted Rand Index (ARI) \cite{ARI}, which quantifies the clustering stability across the 20 repetitions/models, and also guided by prior knowledge in literature. For each model, it assigns each patient to its corresponding subtype membership with the highest probability. The final clustering membership was determined by a consensus clustering strategy across the 20 models. The mapping function ($f$) of Smile-GANs was first applied to visualize the subtypes' atrophy patterns captured for fake PT data generation. Moreover, Voxel-wise group comparison between CN and subtypes of real PT data were performed with AFNI 3dttest \cite{afni} with GM tissue maps \cite{RAVEN}. 

\textbf{Experiments on longitudinal data}
The trained 20 models from ANDI2 was applied to longitudinal data for determining the subtype membership for scans from all visits. Note that only subjects who were consistently assigned into the same subtype from more than 60\% of the models (i.e., 12) were finally taken into account, which were studied for the disease longitudinal progression pathways from prodromal to full AD stages.

\section{Results}
\subsection{Results on simulated data} 
\textbf{WD for training monitoring} Fig. \ref{fig3}(a) shows the change of WD and clustering error (i.e., 1 - clustering accuracy) during training. Generally, the two metrics are consistent in monitoring the training process. Therefore, WD could be used as a surrogate of clustering accuracy when the latter is not available in real applications. Note that inconsistencies at the beginning of training or oscillations exist. Such circumstances can be filtered out by other two metrics, alteration quantity (AQ) and cluster loss. We propose to use WD, along with AQ and cluster loss, as metrics for monitoring the training process. \par
\textbf{Clustering performance comparison across models} Fig. \ref{fig3}(b) shows the clustering accuracy of Smile-GANs, K-means and GMM. Smile-GANs ($0.916\pm0.025$) plainly outperforms the K-means ($0.615\pm 0.028$) and GMM ($0.611\pm0.039$) for clustering those three subtypes. \par 
\textbf{Mapping function for visualization of atrophy patterns} Fig. \ref{fig3}(c) shows the atrophy patterns captured by all different mapping directions (\emph{K}=3). Smile-GANs can automatically identify the ground truth of atrophy patterns, while not capturing any covariate effects. Together with Fig. \ref{fig3}(b), our results clearly indicates that Smile-GANs can not only cluster subtypes accurately, but also automatically identify the underlying atrophy patterns for better interpretation.  

\subsection{Results on ADNI2 Baseline Data}
For baseline experiments, the stopping criteria during training is that the highest WD of all mappings, AQ and cluster loss are smaller than 0.22, 35 and 0.001, respectively. The maximum number of epochs was set to be 6000 and models failing to converge were discarded.\par
\begin{figure}
  \centering
  \includegraphics[width=1\linewidth,height=8.6cm]{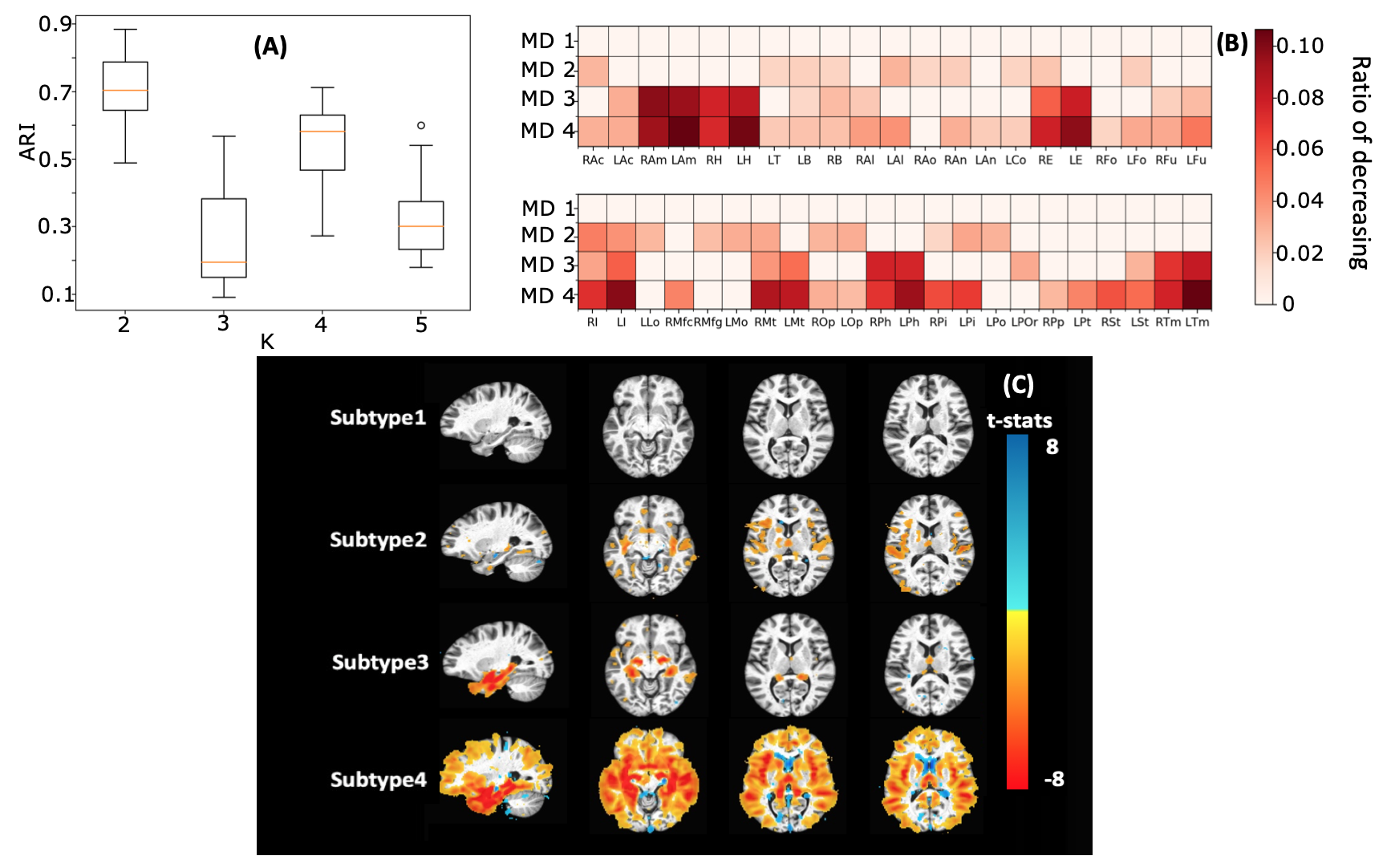}
  \caption{Result on baseline ADNI2. (A): Clustering stability across different \emph{K} with adjusted rand index (ARI). (B): Atrophy patterns of subtypes captured by mapping function (\emph{f}) of Smile-GANs for data generation. Only the ROIs whose values decreasing over 2.5\% are displayed. MD: mapping direction. (C): Voxel-wise statistical comparison between subtype and CN. FDR method with p-value threshold of 0.05 was used for the correction of multiple comparison.}\label{fig4}
\end{figure}

\textbf{Clustering stability for optimal \emph{K}} Fig. \ref{fig4}(A) shows the results of ARI for different \emph{K}.
Though \emph{K}=2 gave the highest ARI, Smile-GANs roughly divided PT into one subtype with barely atrophy and another subtype with whole-brain atrophy. Based on the prior knowledge from the literature \cite{HYDRA}\cite{Chimera} and relatively higher ARI for \emph{K}=4, we therefore selected \emph{K}=4 for following experiments. \par 

\textbf{Neuroanatomical Heterogeneity between Subtypes and CN} Fig. \ref{fig4}(B) shows regions with atrophy identified by the four mapping directions for fake data generation (Detail of the name of ROIs are present in Supplementary). Fig. \ref{fig4}(C) shows the voxel-based group comparison results for each subtype group of real subtypes versus CN group. The two approaches converge to the same findings: i) Subtype 1, referred as normal brain, exhibits no atrophy over the whole brain; ii) Subtype 2, denoted as diffuse atrophy atypical of AD, shows widespread atrophy patterns in frontal and temporal lobe, but medial temporal lobe is spared; iii) Subtype 3, referred as focal medial temporal lobe atrophy, shows localized atrophy in the hippocampus and the anterior-medial temporal cortex; iv) Subtype 4, denoted as typical-AD, displays severe atrophy over the whole brain. \par

\textbf{Clinical Characteristics of clustering subtypes} The clinical characteristics of the four subtypes are summarized in Table \ref{tab1}. Most of subjects in subtype 1 are MCI subjects while more than half of AD subjects are in subtype 4. All three clinical variables (i.e., Abeta, T-tau and WML) of subtype 1 significantly differ to those of the other subtypes. Subtype 2 and 3 also show significant difference in Abeta (p=0.018) and T-tau (p=0.003). Though not significant, subtype 2 has substantially higher WML load than subtype 3.
\begin{table}
  \centering
  \caption{Clinical characteristics of clustering subtypes. WML: white matter lesion}\label{tab1}
  \begin{tabular}{lllll}
    \toprule
    \cmidrule(r){2-5}
        & Subtype 1     & Subtype 2  & Subtype 3& Subtype 4\\
    \midrule
    AD & 5(2.6\%) & 15(15\%)&29(29.6\%)&     89(52.0\%)\\
    MCI & 185(97.4\%)  & 85(85\%)&69(70.4\%) &     82(48.0\%)\\
   Median Abeta & 192.0  & 158.5&142.0&     135.0\\
   Median T-tau    & 62.9  & 74.6&98.1 &    96.0 \\
   Median WML    & 926.4  & 3152.4 &     754.8&11054.9 \\
    \bottomrule
  \end{tabular}
\end{table}
\subsection{Results on Longitudinal Data}
Table \ref{tab2} shows the subtype membership conversion from baseline assignment to future longitudinal assignment using longitudinal AD, MCI and CN subjects. Most of subjects in subtype 1 at baseline remained unchanged for future membership assignment, but some of them progressed to other subtypes in later visits. A substantial proportion of subjects who were assigned into subtype 2 and 3 at baseline finally transformed into subtype 4, but the transformation between these two subtypes are very rare.\par
\begin{figure}
  \centering
  \includegraphics[width=1\linewidth,height=8cm]{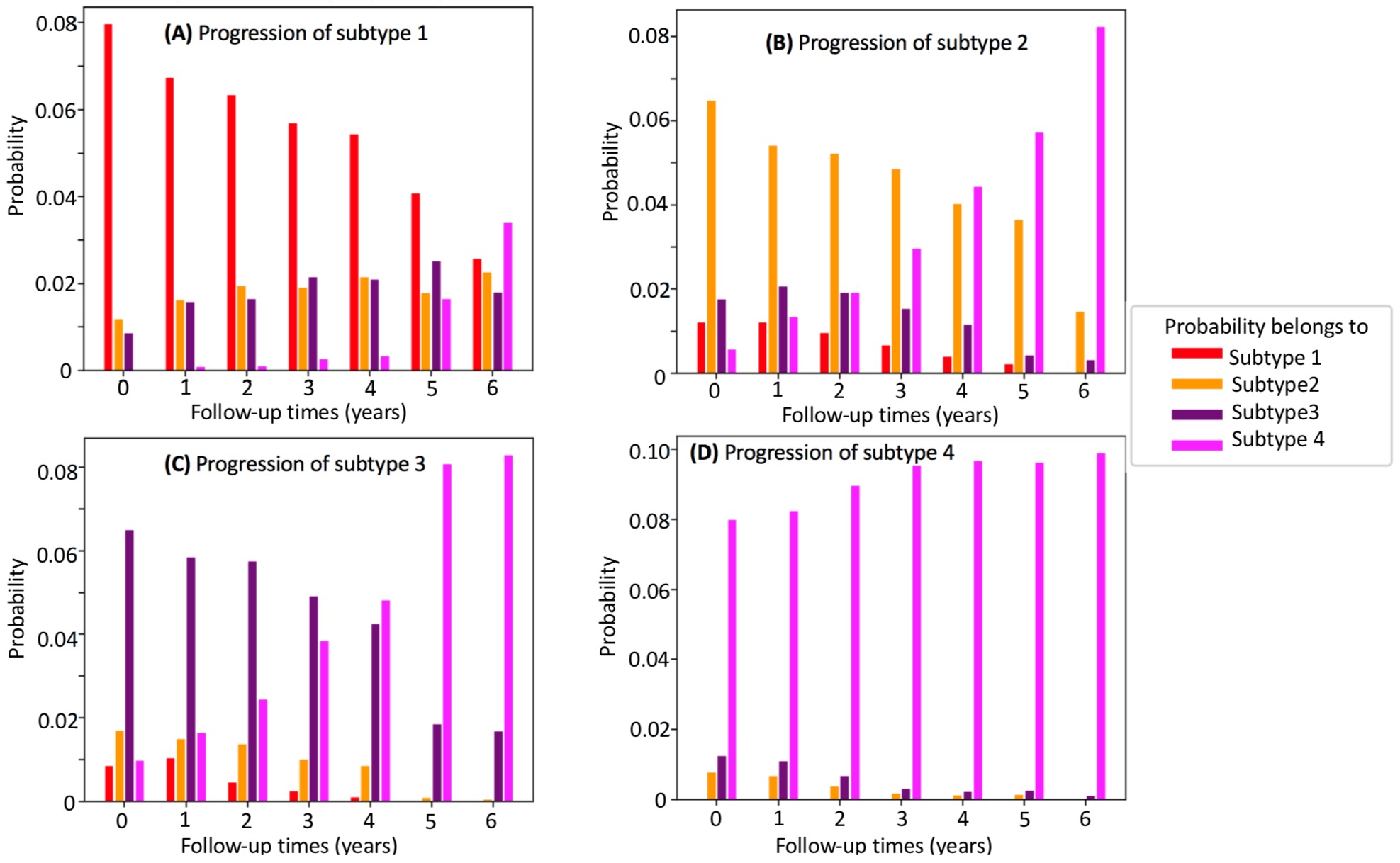}
  \caption{Results on longitudinal conversions of clustering membership. The probability of membership conversion to other subytpes are shown to track the disease progression. Each subject was initially assigned, at its first visit, as subtype 1, 2, 3 or 4, corresponding to (A), (B), (C) and (D) respectively.}\label{fig5}
\end{figure}

\begin{table}
  \centering
  \caption{Results on longitudinal conversions of clustering membership. Subjects from each subtype, determined by the initial clustering membership at their first visit, were reassigned for the membership at their future visits.}\label{tab2}
  \begin{tabular}{lllll}
    \toprule
    \headercell{Initial\\ membership} &\multicolumn{3}{c@{}}{Future membership}                   \\
    \cmidrule(r){2-5}
        & Subtype 1     & Subtype 2  & Subtype 3& Subtype 4\\
    \midrule
    Subgroup 1 & 83.7\%(860/1027)  & 9.3\%(96/1027)&6.23\%(64/1027) &     3.2\%(33/1027)\\
    Subgroup 2    & 3.8\%(16/413)  & 74.8\%(309/413)&2.17\%(9/413) &     20.5\%(85/413) \\
    Subgroup 3    & 4.2\%(11/260)  & 1.9\%(5/260)&62.7\%(163/260) &     32.3\%(84/260) \\
    Subgroup 4    & 0.2\%(1/469)  & 0.8\%(4/469)&0.8\%(4/469) &     98.2\%(460/469)   \\
    \bottomrule
  \end{tabular}
\end{table}
Alternatively, we also studied the change of probability for subtype membership assignment in AD and MCI subjects in a window of 6 years (Fig. \ref{fig5}). For subjects assigned into subtype 1 at baseline, they show increasing probabilities belonging to subtype 2 or 3 at early stage, and then to subtype 4 at later stage (Fig. \ref{fig5} (a)). For subjects of subtype 2 and 3, they all have increasing probabilities belonging to subtype 4 in later visits (Fig. \ref{fig5} (b) and (c)). Those results (Table \ref{tab2} and Fig. \ref{fig5}) potentially indicate two distinct longitudinal disease progression pathways: i) subtypes 1 - 2 - 4, and ii) subtypes 1 - 3 - 4.

\section{Conclusion}
In this study, we proposed a novel method, Smile-GANs, for parsing disease heterogeneity in interpretable ways that support precision diagnostics. Smile-GANs found 4 robust subtypes differing in clinical profiles and unveiled two distinct longitudinal disease progression pathways. Though we demonstrate our claims on the heterogeneity of AD, Smile-GANs is general and able to be applied to other medical applications and domains. The direction of our future work is to extend the current model to high-dimensional imaging data (i.e., voxel-wise features) to better capture multivariate patterns of disease heterogeneity. \par

\section*{Broader Impact}
The current work has the following potential positive impacts. First, Simle-GANs provides a general and principled way of capturing biological heterogeneity in an interpretable way, hence it can help in mode precisely defining many diseases and pathologies, based on quantitative measures like imaging. Herein we present AD as an example, and demonstrate how dissecting heterogeneity of this disease and its prodromal stage can help us identify different paths to dementia.  Second, the current work potentially provides some reasoning for the failure of disease-modifying treatments in AD, which might be more effective if applied to the right patient subpopulations. The breakdown of disease heterogeneity enables to refine relatively unique pathologic populations and eventually benefit future therapeutic trials. Meanwhile, it should be mentioned that any predictive machine learning model runs risk of misclassification, our work is no exception. However, our model primarily seeks to stratify the patient population, avoiding the detrimental false positive screening results for participants.


\newpage

\section{Supplementary}
\subsection{Network Architecture}
Detailed architectures of mapping function, clustering function and discriminator are provided in table \ref{tab4} and table \ref{tab5}.
\begin{table}[H]
    \centering
     \caption{Architecture of mapping function $f$}\label{tab4}
    \begin{tabular}{cccccc}
    \toprule
    &Layer& Input Size  & Bias Term &  leaky relu $\alpha$& Output Size\\ 
    \midrule
    Phase1(Encoder)&
    \begin{tabular}{c}
    Linear1+Leaky-Relu\\Linear2+Leaky-Relu
    \end{tabular}&
    \begin{tabular}{c}
    145*1\\72*1
    \end{tabular}&
    \begin{tabular}{c}
    No\\No
    \end{tabular}&
    \begin{tabular}{c}
    0.2\\0.2
    \end{tabular}&
    \begin{tabular}{c}
    72*1\\36*1
    \end{tabular}\\
    \midrule
    Phase1(Decoder)& Linear1+Sigmoid & K*1& Yes & NA &36*1\\
    \midrule
    Phase2&
    \begin{tabular}{c}
    Linear1+Leaky-Relu\\Linear2+Leaky-Relu\\Linear3
    \end{tabular}&
    \begin{tabular}{c}
    36*1\\72*1\\145*1
    \end{tabular}&
    \begin{tabular}{c}
    No\\No\\No
    \end{tabular}&
    \begin{tabular}{c}
    0.2\\0.2\\NA
    \end{tabular}&
    \begin{tabular}{c}
    72*1\\145*1\\145*1
    \end{tabular}\\
    \bottomrule
    \end{tabular}
\end{table}

\begin{table}[H]
    \centering
    \caption{Architecture of discriminator $D$ and clustering function $g$}\label{tab5}
    \begin{tabular}{cccccc}
    \toprule
    &Layer& Input Size  & Bias Term &  leaky relu $\alpha$& Output Size\\ 
    \midrule
    Discriminator&
    \begin{tabular}{c}
    Linear1+Leaky-Relu\\Linear2+Leaky-Relu\\Linear3+Softmax
    \end{tabular}&
    \begin{tabular}{c}
    145*1\\72*1\\36*1
    \end{tabular}&
    \begin{tabular}{c}
    Yes\\Yes\\Yes
    \end{tabular}&
    \begin{tabular}{c}
    0.2\\0.2\\NA
    \end{tabular}&
    \begin{tabular}{c}
    72*1\\36*1\\2*1
    \end{tabular}\\
    \midrule
    Clustering&
    \begin{tabular}{c}
    Linear1+Leaky-Relu\\Linear2+Leaky-Relu\\Linear3+Leaky-Relu\\Linear4+Softmax
    \end{tabular}&
    \begin{tabular}{c}
    145*1\\145*1\\72*1\\36*1
    \end{tabular}&
    \begin{tabular}{c}
    Yes\\Yes\\Yes\\Yes
    \end{tabular}&
    \begin{tabular}{c}
    0.2\\0.2\\0.2\\NA
    \end{tabular}&
    \begin{tabular}{c}
    145*1\\72*1\\36*1\\K*1
    \end{tabular}\\
    \bottomrule
    \end{tabular}
\end{table}

\subsection{Algorithm}
Detailed training procedure of Smile-GANs is disclosed by Algorithm \ref{algo}.
\IncMargin{1em}
\begin{algorithm}
\caption{Smile-GANs training procedure. $l_c$ represents cross entropy loss and $e_i$ represents a one hot vector with 1 at $i_{th}$ component.}\label{algo}
\SetKwData{Left}{left}\SetKwData{This}{this}\SetKwData{Up}{up}
\SetKwFunction{Union}{Union}\SetKwFunction{FindCompress}{FindCompress}
\SetKwInOut{Input}{input}\SetKwInOut{Output}{output}
\BlankLine
\While{not meeting stopping criteria or reaching max\_epoch}{
\For{all batches $\{x_i\}_{i=1}^m$,$\{y_i\}_{i=1}^m$}{
\emph{sample m integers $\{a_i\}_{i=1}^m$ with $a_i\sim \text{discrete-}U(1,K)$ and let $z_i=e_{a_i}$}
\bigbreak
\textbf{Update weights of discriminator D:}
Use ADAM to update $\theta_D$ with gradient:
$\nabla_{\theta_D}\frac{1}{m}\sum_{i=i}^m[(l_c(D(y_i),e_1)+l_c(D(f(x_i,z_i),e_0)))]$
\bigbreak
\textbf{Update weights of mapping function f:}
Use ADAM to update $\theta_f$ with gradient:
$\nabla_{\theta_f}\frac{1}{m}\sum_{i=i}^m[l_c(D(f(x_i,z_i),e_1)))+\lambda l_c(g(f(x_i,z_i),z_i)+\mu(||f(x_i,z_i)-x_i||_1)]$
\bigbreak
\textbf{Update weights of clustering function g:}
Use ADAM to update $\theta_g$ with gradient:
$\nabla_{\theta_g}\frac{1}{m}\sum_{i=i}^m[\lambda l_c(g(f(x_i,z_i),z_i)+\mu(||f(x_i,z_i)-x_i||_1)]$
}
}
\end{algorithm}\DecMargin{1em}

\subsection{ROIs names}
Full names of ROIs shown in Fig. 3(B) are displayed in table \ref{tab6}.
\begin{table}[H]
  \centering
  \caption{Full names of ROIs}\label{tab6}
  \begin{tabular}{llll}
  \toprule
    Abbr     & ROI  & Abbr & ROI\\
    \midrule
    RAC & Right Accumbens Area  & RI&Right Inferior Temporal Gyrus \\
    LAC & Left Accumbens Area  & LI &Left Inferior Temporal Gyrus \\
    RAm & Right Amygdala  & LLO& Left Lateral Orbital Gyrus \\
    LAm & Left Amygdala  & RMfc&Right Medial Frontal Cortex \\
    RH & Right Hippocampus  & RMfg&Right Middle Frontal Gyrus \\
    LH & Left Hippocampus  & LMo&Left Middle Occipital Gyrus \\
    LT & Left Thalamus Proper  & RMt&Right Middle Temporal Gyrus \\
    LB & Left Basal Forebrain  & LMt&Left Middle Temporal Gyrus \\
    RB & Right Basal Forebrain  & ROp&Right Opercular Part of the Inferior Frontal Gyrus \\
    RAI & Right Anterior Insula  & LOp&Left Opercular Part of the Inferior Frontal Gyrus \\
    LAI & Left Anterior Insula  & RPh&Right Parahippocampal Gyrus\\
    RAO & Right Anterior Orbital Gyrus  & LPh&Left Parahippocampal Gyrus \\
    RAn & Right Angular Gyrus  & RPi&Right Posterior Insula\\
    LAn & Left Angular Gyrus  & LPi&Left Posterior Insula \\
    LCo & Left Central Operculum  & LPo&Left Parietal Operculum \\
    RE & Right Entorhinal Area  & LPOr&Left Posterior Orbital Gyrus \\
    LE & Left Entorhinal Area  & RPp&Right Planum Polare \\
    RFo & Right Frontal Operculum  & LPt&Left Planum Temporale \\
    LFo & Left Frontal Operculum  & RSt&Right Superior Temporal Gyrus \\
    RFu & Right Fusiform Gyrus  & LSt&Left Superior Temporal Gyrus \\
    LFu & Left Fusiform Gyrus  & RTm&Right Temporal Pole \\
     &                   & LTm&Left Temporal Pole \\
    \bottomrule
  \end{tabular}
\end{table}

\end{document}